\documentclass[aps,reprint,article,prl]{revtex4-1}
\usepackage{amsmath}
\usepackage{dcolumn}
\usepackage{graphicx}
\usepackage{float}

\begin{document}
\title{Friedel Oscillations of Vortex Bound States Under Extreme Quantum Limit in KCa$_2$Fe$_4$As$_4$F$_2$}
\author{Xiaoyu Chen,$^{1,*}$ Wen Duan,$^{1,*}$ Xinwei Fan,$^{1,*}$ Wenshan Hong,$^{2,3,*}$ Kailun Chen,$^1$ Huan Yang,$^{1,\dag}$ Shiliang Li,$^{2,3,4}$ Huiqian Luo,$^{2,4,\ddagger}$ and Hai-Hu Wen$^{1,\S}$}

\affiliation{National Laboratory of Solid State Microstructures and Department of Physics, Collaborative Innovation Center of Advanced Microstructures, Nanjing University, Nanjing 210093, China}

\affiliation{Beijing National Laboratory for Condensed Matter Physics, Institute of Physics, Chinese Academy of Sciences, Beijing 100190, China}

\affiliation{School of Physical Sciences, University of Chinese Academy of Sciences, Beijing 100190, China}

\affiliation{Songshan Lake Materials Laboratory, Dongguan, Guangdong 523808, China}

\begin{abstract}
We report the observation of discrete vortex bound states with the energy levels deviating from the widely believed ratio of $1:3:5$ in the vortices of an iron based superconductor KCa$_2$Fe$_4$As$_4$F$_2$ through scanning tunneling microcopy (STM). Meanwhile Friedel oscillations of vortex bound states are also observed for the first time in related vortices. By doing self-consistent calculations of Bogoliubov-de Gennes equations, we find that at extreme quantum limit, the superconducting order parameter exhibits a Friedel-like oscillation, which modifies the energy levels of the vortex bound states and explains why it deviates from the ratio of $1:3:5$. The observed Friedel oscillations of the bound states can also be roughly interpreted by the theoretical calculations, however some features at high energies could not be explained. We attribute this discrepancy to the high energy bound states with the influence of nearby impurities. Our combined STM measurement and the self-consistent calculations illustrate a generalized feature of the vortex bound states in type-II superconductors.
\end{abstract}

\maketitle

For a vortex in a type-II superconductor, it is generally understood that the quantized magnetic flux of $\Phi_0 = h/2e = 2.07\times10^{15}$ Wb distributes in the region with the radius of penetration depth $\lambda$; while the order parameter $\psi$ ramps up from zero at the core center to the full value $\psi_0$ in the scale characterized by the coherence length $\xi$. The vortex core region which has some features of the normal-state can be regarded as a kind of quantum well surrounded by the gapped superconducting region, and naturally the vortex bound states (VBS) can appear based on the solutions to the Bogoliubov-de-Gennes (BdG) euqations \cite{CdGM,Kramer Pesh,Klevin}. An early simplified analytic solution to the BdG equations indicates that the discrete energy levels of bound states should appear \cite{CdGM} at $E_\mu = \pm\mu\Delta^2/E_\mathrm{F}$, and it was argued later that $\mu$ = 1/2, 3/2, 5/2, $\cdots$ \cite{Kramer Pesh}, with $\Delta$ the superconducting gap and $E_\mathrm{F}$ the Fermi energy. Later on, the quantized VBS were predicted based on the self-consistent calculations of the BdG equations \cite{selfconsistent1,Berthod}, which also yields the spatial dependence of the superconducting gap $\Delta(r)\propto \tanh(r/\xi_0)$; here, $\xi_0=v_\mathrm{F}/\Delta_0$ with $v_\mathrm{F}$ the Fermi velocity and $\Delta_0$ the superconducting gap at $T=0$. In most superconductors, we have $E_\mathrm{F}\gg\Delta$, which is the basic requirement of the BCS theory in the weak coupling limit. Therefore, the energy spacing for neighboring bound states $\Delta^2/E_\mathrm{F}$ is too small to be discernible; what was shown by experimental observations is that a particle-hole symmetric VBS peak (assembled by many crowded bound states) locates at zero energy and then it splits and fans out when moving away from the vortex center \cite{selfconsistent1,Berthod,Hess1,Hess2,vortex review}.

The discrete bound states can be however distinguished when the thermal smearing energy is smaller than the energy spacing of VBS \cite{selfconsistent2}, i.e., in the quantum limit of $T/T_\mathrm{c}\leq\Delta/E_\mathrm{F}$ or $T/T_\mathrm{c}\leq1/(k_\mathrm{F}\xi_0)$ with $T_\mathrm{c}$ the critical temperature and $k_\mathrm{F}$ the Fermi wave vector. This can be achieved in superconductors with a relatively large value of $\Delta/E_\mathrm{F}$. Furthermore, under the extreme quantum limit (EQL),  $T/T_\mathrm{c}\ll\Delta/E_\mathrm{F}$, it was shown that the $\Delta(r)$ should exhibit an oscillatory spatial variation with a period of about $\pi/k_\mathrm{F}$ instead of the monotonic variation behavior of $\tanh(r/\xi_0)$ \cite{selfconsistent2}, and meanwhile Friedel oscillations of the charge profile which is related to the local density of states (LDOS) were predicted \cite{selfconsistent2,Friedel charge1,Friedel charge2}. Experimentally, some traces of discrete VBS were reported as an asymmetric peak or two close peaks near zero-bias measured in vortex centers of cuprates \cite{STMReview}, YNi$_2$B$_2$C \cite{YNiBC}, and some iron-based superconductors \cite{BaKFeAs,LiFeAs,HoffmanReview,Suderow1144,HanaguriFeSe}. It was shown later that the two close peaks in cuprates are not the VBS \cite{Berthod1,Berthod2}. Recently, discrete VBS were clearly observed in vortices measured in FeTe$_{0.55}$Se$_{0.45}$ with the peak-energy ratio near $1:3:5$ \cite{CdGM NC}. Clear discrete bound state peaks were also observed in the FeSe monolayer \cite{FeSe/SrTiO3}, and they were observed coexisting with the Majorana zero mode in FeTe$_{0.55}$Se$_{0.45}$ \cite{FTS DH}, (Li$_{0.84}$Fe$_{0.16}$)OHFeSe \cite{LiFeOHFeSe}, and KCaFe$_4$As$_4$ \cite{KCaFe4As4}. Some preliminary signatures of Friedel oscillations of the bound states were also reported around vortices in some iron based superconductors \cite{CdGM NC,LiFeAs,HanaguriFeSe}, these were discussed as the consequence of VBS in the quantum limit.

In this Letter, we present scanning tunneling microscopy/spectroscopy (STM/STS) results of the VBS in KCa$_2$Fe$_4$As$_4$F$_2$. Clear discrete VBS are observed with energy ratio deviating from the expected $1:3:5$, and Friedel oscillations are also observed in some vortices as the VBS in the extreme quantum limit (EQL). We also observe some Friedel oscillations surrounding the vortex center at energies near the superconducting gap, which cannot be explained by the theory in the clean limit. Combining with the theoretical calculations, we conclude that the explicit evidence of the behaviors for the VBS under EQL have been found.

KCa$_2$Fe$_4$As$_4$F$_2$ (K12442) is a newly found iron based superconductor with $T_\mathrm{c}=33.5$ K \cite{K12442}. Several Fermi surfaces with multiple and nodeless gaps were observed by the angle-resolved photoemission spectroscopy (ARPES) \cite{ARPES}, and some of the bands are very shallow. It is found that the dominant contribution of density of states arises from the shallow $\alpha$ pocket based on STM/STS measurements \cite{DuanSTM}. Hence, the VBS should be interesting in this multiple and shallow band superconductor. Single crystals of KCa$_2$Fe$_4$As$_4$F$_2$ were grown by the self-flux method \cite{12442Hc2}. STM/STS measurements were carried out in a scanning tunneling microscope (USM-1300, Unisoku Co., Ltd.). K12442 samples were cleaved at about 77 K in an ultrahigh vacuum with the base pressure of about $1\times10^{-10}$ Torr, and then they were transferred to the microscopy head which was kept at a low temperature. Electrochemically etched tungsten tips were used for STM/STS measurements after the electron-beam heating. A typical lock-in technique was used in STS measurements with an ac modulation of 0.1 mV and the frequency of 931.773 Hz. Setpoint conditions are $V_\mathrm{set} = 10$ mV and $I_\mathrm{set} = 200$ pA; temperature is 0.4 K for all STM/STS measurements.

\begin{figure}
\centering
\includegraphics[width=8.6cm]{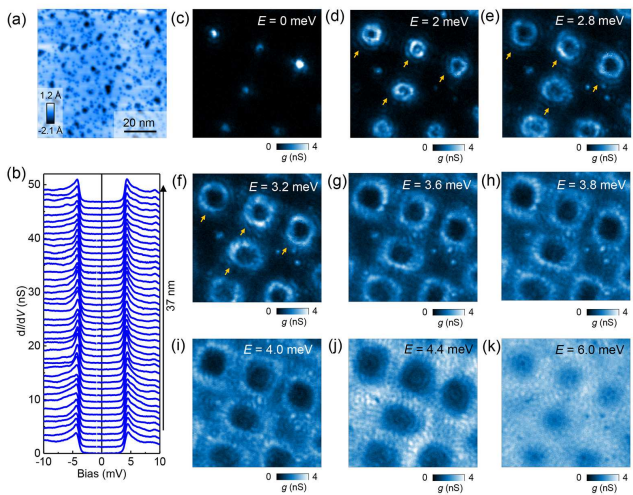}
\caption{(a) Topographic image measured in KCa$_2$Fe$_4$As$_4$F$_2$. (b) A set of tunneling spectra measured under $\mu_0H = 0$ T and along a line with the length of 37 nm in the region near the shown area of (a). (c)-(k) Vortex images acquired at different biases and under $\mu_0H = 2$ T; the mapping area is the same as the one shown in (a).
} \label{fig1}
\end{figure}

Figure~\ref{fig1}(a) shows a typical topography which is commonly obtained on the cleaved surface of K12442 \cite{DuanSTM}. The flat area is the $\sqrt{2}\times\sqrt{2}$ reconstructed surface by half amount of K or Ca atoms of the related layer after the cleavage, while there are many hollows with different sizes on the surface which may be the assembled vacancies \cite{DuanSTM}. However, the tunneling spectra measured in this kind of area show very homogeneous feature with coherence peak energies at about $4.3\pm0.2$ meV, and Fig.~\ref{fig1}(b) shows a set of tunneling spectra as an example. These spectra are measured along a line with the length of 37 nm near the displayed area of Fig.~\ref{fig1}(a). Then we try to image vortices under the magnetic field of 2 T, and Figs.~\ref{fig1}(c)-\ref{fig1}(k) show vortex images by the spatially resolved differential conductance (d$I$/d$V$). Although the scanning area is not big enough to see many vortices, one can still see that vortices are randomly but almost equidistant distributed. The expected hexagonal or square vortex lattice do not show up in the mapping area, which suggests that the vortex pinning may be strong in the sample. In Fig.~\ref{fig1}(c), we show an image of differential conductance measured at $E=0$ meV, and one can see vortex cores with the diameters of about $\xi\approx 1.4$-3 nm. Here the core size is derived from the half-width of maximum of the spatial dependence of LDOS. This value is roughly consistent with the coherence length of about 1.2 \cite{12442Hc2} or 2.2 nm \cite{Hc2HighT} derived from transport measurements by using different criterions. With increase of the bias voltage, the vortex pattern changes from a bright spot to a bright ring, and the diameter of the bright ring increases even to about 18 nm at $E=3.6$ meV [Fig.~\ref{fig1}(g)]. Such bright rings are explained as the spatial evolution of the VBS at finite in-gap energies \cite{LiFeAs,FTS DH,KCaFe4As4}. When the energy exceeds $\Delta\approx4.3$ meV, the structure of the bright ring disappears, with only the dark-disc feature left in the vortex core. However, being different from the structureless dark disc observed in superconducting Bi$_2$Te$_3$/FeTe$_{0.55}$Se$_{0.45}$ heterostructures \cite{ChenSA}, the dark disc here shows some internal structure in vortex images mapped at energies near $\Delta$ in K12442 [Figs.~\ref{fig1}(h)-\ref{fig1}(j)].

\begin{figure}
\centering
\includegraphics[width=8.6cm]{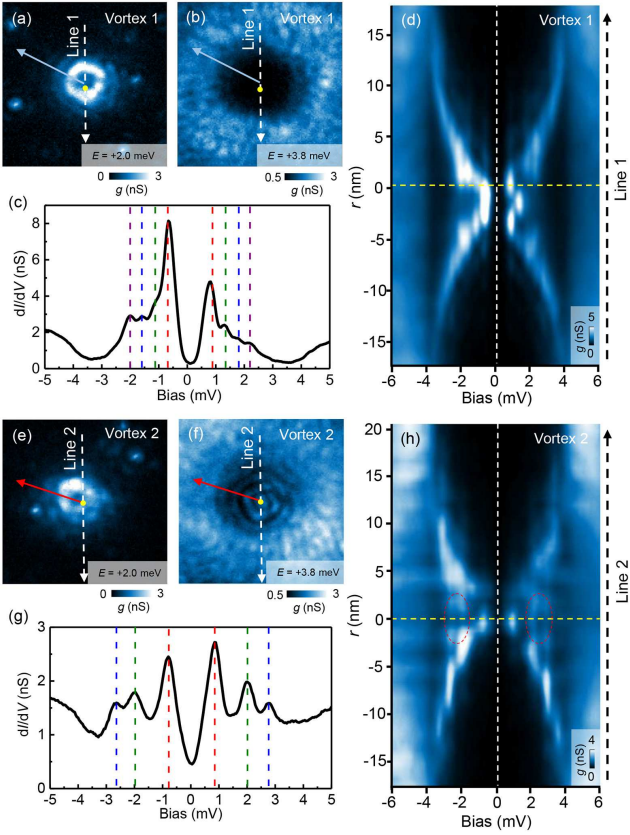}
\caption{(a,b),(e,f) Vortex images measured at different energies and under $\mu_0H = 0.2$ T. (c),(g) Tunneling spectra measured at the yellow dots in (b) and (f), respectively. (d),(h) The line profiles of tunneling spectra across vortex centers for vortex 1 and 2, respectively. Friedel oscillations can be observed in the dark-disc region of vortex 2 at +3.8 meV (f), but they are absent in vortex 1 (b).
} \label{fig2}
\end{figure}

The mapping of a single vortex is carried out under a small magnetic field of 0.2 T in order to minimize the vortex-vortex interaction. Figure~\ref{fig2} shows vortex images and tunneling spectra measured in two typical vortices without (vortex 1, Fig.~\ref{fig2}(b)) and with (vortex 2, Fig.~\ref{fig2}(f)) spatial oscillations in dark disc regions at $E=+3.8$ meV. Tunneling spectra measured at centers of the two selected vortices show clear in-gap bound states [Figs.~\ref{fig2}(c) and \ref{fig2}(g)]. The energies of bound state peaks are almost spatially independent. The higher order bound state with larger peak energy shows up when the position is far away from the vortex center [Figs.~\ref{fig2}(d) and \ref{fig2}(h)]. These features are consistent with the expectation of VBS in the EQL. In addition, two sets of bright rings with weakened brightness can be clearly seen outside the bright ring in vortex 1 [Fig.~\ref{fig2}(a)], and such structures can also been seen around the vortices as indicated by arrows in Figs.~\ref{fig1}(d)-\ref{fig1}(f). This kind of Friedel oscillations is consistent with the theoretical prediction that the peak amplitude for a selected bound state will show spatial oscillation in the EQL \cite{selfconsistent2,YNiBC}. Here, the bound-state energies are about $0.8$, $1.3$, $1.8$ and $2.2$ meV for vortex 1 [Fig.~\ref{fig2}(c)], and they are about $0.9$, $2.1$, and $2.8$ meV for vortex 2 [Fig.~\ref{fig2}(g)]. Therefore, the corresponding ratio of bound state energies is $1:1.6:2.3:2.8$ and $1:2.3:3.1$ for vortex 1 and 2, respectively; both of these values are deviating from the ratio of $E_{1/2}:E_{3/2}:E_{5/2}=1:3:5$ \cite{Kramer Pesh}.

\begin{figure}
\centering
\includegraphics[width=8.6cm]{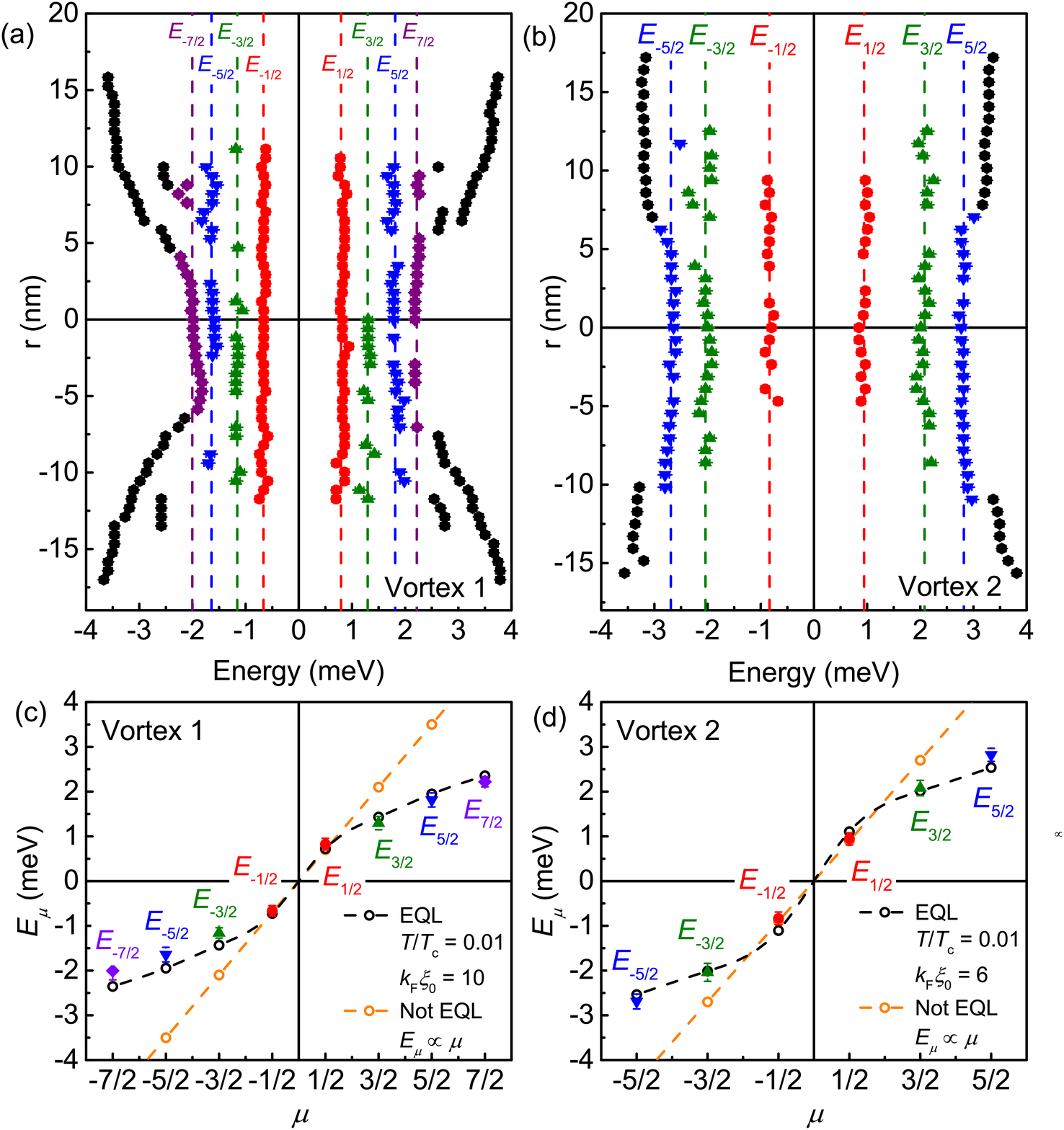}
\caption{(a),(b) Spatial variations of the bound-state energy derived from the tunneling spectra shown in Figs.~\ref{fig2}(d) and \ref{fig2}(h) for vortex 1 and vortex 2, respectively. Peaks with similar energies are indexed by the same order of the VBS. The dark filled symbols represent the peak positions of many assembled VBS at high energies, thus the energy is not fixed. (c),(d) Averaged bound-state energy derived from experiments (full symbols), and theoretical results (open symbols) of bound state energy calculated under and not under the quantum limit.
} \label{fig3}
\end{figure}

Since there are many orders of VBS in line profiles of tunneling spectra across vortex centers shown in Figs.~\ref{fig2}(d) and \ref{fig2}(h), the peak energies of these bound states can be extracted and they are plotted in Figs.~\ref{fig3}(a) and \ref{fig3}(b). By ascribing the bound state peaks with similar energies to the same order of the VBS, we can derive averaged values of bound state energies and show them in Figs.~\ref{fig3}(c) and \ref{fig3}(d) for the two selected vortices. Obviously, the averaged bound-state energy is deviating from the theoretical relationship of $E_\mu = \mu\Delta^2/E_\mathrm{F}$. It should be noted that the bound state energy does not have to follow this relation at extreme low temperatures, namely in the EQL of $T/T_\mathrm{c}\ll\Delta/E_\mathrm{F}$, when there are some oscillations in $\Delta(r)$ which makes the bound-state energy deviate from the linear relation $E_\mu = \mu\Delta^2/E_\mathrm{F}$ \cite{selfconsistent2}. Then we do self-consistent calculations of BdG equations based on the routes in previous reports \cite{selfconsistent1,selfconsistent2}. The detailed calculations will be presented separately.
One can see in Figs.~\ref{fig3}(c) and \ref{fig3}(d) that the theoretical curves of $E_\mu$ match our experimental data well with different values of $k_F\xi_0$ at the limit of $T/T_\mathrm{c} = 0.01$. It should be noted that $\xi/\xi_0\approx(k_\mathrm{F}\xi_0)^{-1}$ in quantum limit \cite{selfconsistent2}, so here $\xi_0$ used in our calculations is much larger than the core size $\xi$ determined from the experiments.

\begin{figure}
\centering
\includegraphics[width=8.6cm]{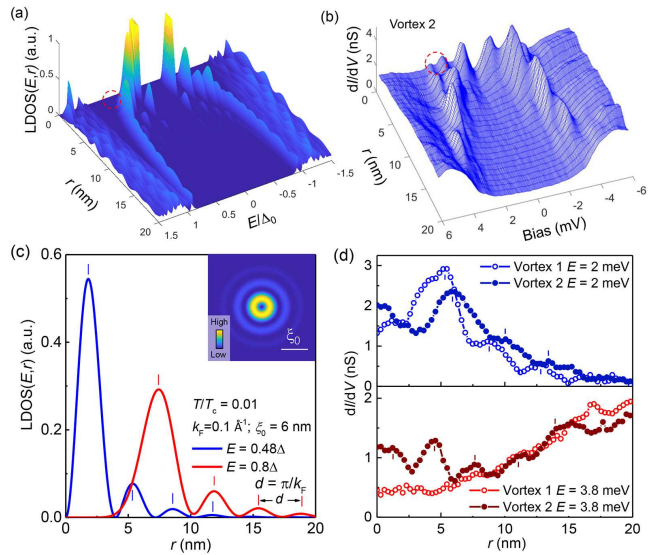}
\caption{(a) Line profile of local DOS across a vortex core from theoretical calculations under the EQL ($T/T_\mathrm{c} = 0.01$, $k_F\xi_0=6$). (b) Three-dimensional plot of the partial of tunneling spectra shown in Fig.~\ref{fig2}(h). (c) Theoretical results of the spatial variation of LDOS at two selected energies. The inset shows the two-dimensional color plot of the LDOS of a vortex core calculated at $E=0.48\Delta$. (d) Experimental results of the spatial variation of differential conductance measured across two vortex cores along the blue and red arrowed line in Figs.~\ref{fig2}(a,b) and ~\ref{fig2}(e,f), respectively.
} \label{fig4}
\end{figure}

Based on self-consistent calculations of BdG equations, the line profile of LDOS across a vortex core under the EQL is shown in Fig.~\ref{fig4}(a). The result clearly shows discrete bound state peaks and spatial oscillations of LDOS. These Friedel oscillations can be clearly seen at fixed energies, two typical examples are given in Fig.~\ref{fig4}(c) for different energies. For the low energy one at $E = 0.48\Delta$, the first-order oscillation appears at a small distance away from the center, which is followed by several sets of oscillations with decaying intensities. The two dimensional mapping of the calculated LDOS for this energy is plotted as an inset to Fig.~\ref{fig4}(c). This calculation result is roughly consistent with the experimental observations as shown in Fig.~\ref{fig2}(a) and 2(e) for both vortex 1 and vortex 2.  At a high energy of  $E = 0.8\Delta$, the calculated LDOS will be expelled from the core region close to the center, and the first-order oscillation appears at a distance of about 7-8 nm away from the center. And the higher orders of oscillations appear outside the first-order ring with decaying intensities when moving outside. Fig.~\ref{fig4}(b) shows three-dimensional plot of the measured LDOS for vortex 2 in the positive half space, which mimics the calculation result in Fig.~\ref{fig4}(a). The measured differential conductance at roughly similar corresponding energies are presented in Fig.~\ref{fig4}(d). One can see that for vortex 1 at $E = 3.8$ meV, the intensity of LDOS is indeed strongly suppressed in the core region and a weak first-order oscillation appears at a distance away from the center. Thus we can conclude that the experimental data are roughly consistent with calculation results for vortex 1, and also for vortex 2 at a low energy, see open symbols and filled blue circles in Fig.~\ref{fig4}(d), respectively. But the situation at a high energy seems to be different for vortex 2 in which the Friedel oscillations can be seen very close to the vortex center (now $E=+3.8$ meV), see Fig.~\ref{fig2}(f) and the purplish-red filled circles in Fig.~\ref{fig4}(d). In addition, the lower-order bound-state peaks ($E_{\pm3/2}$ and $E_{\pm5/2}$) also have spatial oscillations as marked by the red circle in Fig.~\ref{fig4}(b). But interestingly, even with these different features in vortex 1 and 2, the oscillation periodicity seems to be similar to each other. Since Friedel oscillations should have the periodicity\cite{selfconsistent2} of about $\pi/k_\mathrm{F}$, we just calculate the value no matter where the Friedel oscillations locate in these two vortices. The obtained period of $\pi/k_\mathrm{F}$ is about 3.6 nm for vortex 1 and 3.0 nm for vortex 2.

In the vortices measured on the K12442 sample, discrete VBS are observed with the energy levels deviating from the ratio of $1:3:5$. This is due to a relatively high $T_\mathrm{c}$ and large value of $\Delta/E_\mathrm{F}$, which makes the extreme quantum limit condition $T/T_\mathrm{c}\ll\Delta/E_\mathrm{F}$ easily satisfied. The Friedel oscillation can be clearly seen around the vortices also because the EQL is satisfied. The smaller $k_\mathrm{F}$ or $k_\mathrm{F}\xi_0$ makes the relation of $E_\mu$ more deviating from the linear relation of $E_\mu=\mu\Delta^2/E_\mathrm{F}$ although the quantum limit condition $T/T_\mathrm{c}\leq\Delta/E_\mathrm{F}$ are all satisfied \cite{Friedel charge1}. In K12442, the dominant scattering is the intra-band scattering of the hole-like $\alpha$ pocket which has a small $k_\mathrm{F}$. As mentioned above, the feature of VBS in vortex 2 at a high energy is not compatible with current theoretical calculations. In Fig.~\ref{fig4}(d), one can see that the background differential conductance at the vortex center is much higher for vortex 2 when compared with the one for vortex 1. A reasonable explanation is that the impurity scattering is strong in core area of vortex 2, which will bring in more complex into the calculation to the BdG equations. Thus we believe that the discrepancy between the features of vortex 2 at a high energy and the related theoretical calculations is induced by the effect of impurities in the core region, which modifies the total Hamiltonian involved in the calculations. Unfortunately theoretical considerations on the VBS due to both the vortex confinement and the impurity effect are still lacking and thus highly desired. Our results of the bound state energies deviating from the ratio of $1:3:5$ and the observations of the Friedel oscillations clearly indicate that the extreme quantum limit condition is satisfied in present system.

In conclusion, we have observed discrete vortex bound states with energies deviating from $1:3:5$ in KCa$_2$Fe$_4$As$_4$F$_2$. Friedel oscillations of the vortex bound states are also observed. These two unique features are consistent with our self-consistent calculations on the BdG equations under the extreme quantum limit. However, in some vortices at energies close to the gap, we observe the Friedel oscillations staring from the vortex core center, this cannot be explained by the present theoretical frame. We attribute this discrepancy to the cooperative effect by both the vortex confinement and impurity scattering. Our results inspire a more complete theoretical treatment to include also the impurity scattering when solving the BdG equations, and should shed new light on a generalized understanding on the vortex core state in a type-II superconductor.

We appreciate very useful discussions with Christophe Berthod and Da Wang. This work was supported by National Key R\&D Program of China (Grants No. 2016YFA0300401, No. 2018YFA0704200, No. 2017YFA0303100, and No. 2017YFA0302900), National Natural Science Foundation of China (Grants No. 12061131001, No. 11974171, No. 11822411, No. 11961160699, No. 11674406, and No. 11674372), and the Strategic Priority Research Program (B) of Chinese Academy of Sciences (Grants No. XDB25000000, and No. XDB33000000). H. L. is grateful for the support from Beijing Natural Science Foundation (Grant No. JQ19002) and the Youth Innovation Promotion Association of CAS (Grant No. 2016004).

$^*$ These authors contributed equally to this work.

Corresponding authors:

$^\dag$ huanyang@nju.edu.cn

$^\ddag$ hqluo@iphy.ac.cn

$^\S$ hhwen@nju.edu.cn

\end{document}